\documentclass[11pt]{article}
\usepackage{moriond,epsfig}

\def\Journal#1#2#3#4{{#1} {\bf #2}, #3 (#4)}

\def\NIM{\em Nucl. Instrum. Methods}

\def\PLB{{\em Phys. Lett.}  B}
\def\PRL{\em Phys. Rev. Lett.}
\def\PRD{{\em Phys. Rev.} D}

\def\D0{$\rm D\emptyset$}

\def\ra{\rightarrow}

\def\be{\begin{equation}}
\def\ee{\end{equation}}
\def\bea{\begin{eqnarray}}
\def\eea{\end{eqnarray}}

\begin{document}
\vspace*{4cm}
\title{Charm Physics at the Tevatron}

\author{A. Korn}

\address{Laboratory for Nuclear Science,\\ 
Massachusetts Institute of Technology\\
77 Massachusetts Avenue, Cambridge, MA 02139\\
E-mail: akorn@fnal.gov\\
(representing the CDF and \D0 collaboration)}

\maketitle\abstracts{First charm physics results from the CDF and \D0
experiments at the Tevatron Run II are presented. With the addition of
the Secondary Vertex Trigger CDF has become a competitive charm experiment.}

\section{Introduction}

The cross section of $p\bar{p}$ into charm is very high compared 
to $e^+e^-$-machines, but it is orders of magnitude smaller than
the total cross section of $\sim 100 \rm mb$.
This explains the need for a good trigger mechanism. Traditionally 
charm physics at hadron colliders relies on a lepton signature. 
For example, the decay of the $\rm J/\psi$  into two leptons or 
semi-leptonic decays of D-mesons.
 
Both detectors at the Tevatron, CDF and \D0 have undergone substantial
upgrades for RUN II~\cite{TDR}. CDF now exploits a new trigger 
technique selecting more abundant hadronic decays.

\section{$\boldmath \rm J/\psi$ Cross Section}

\begin{figure}[!h]
\begin{center}
\psfig{figure=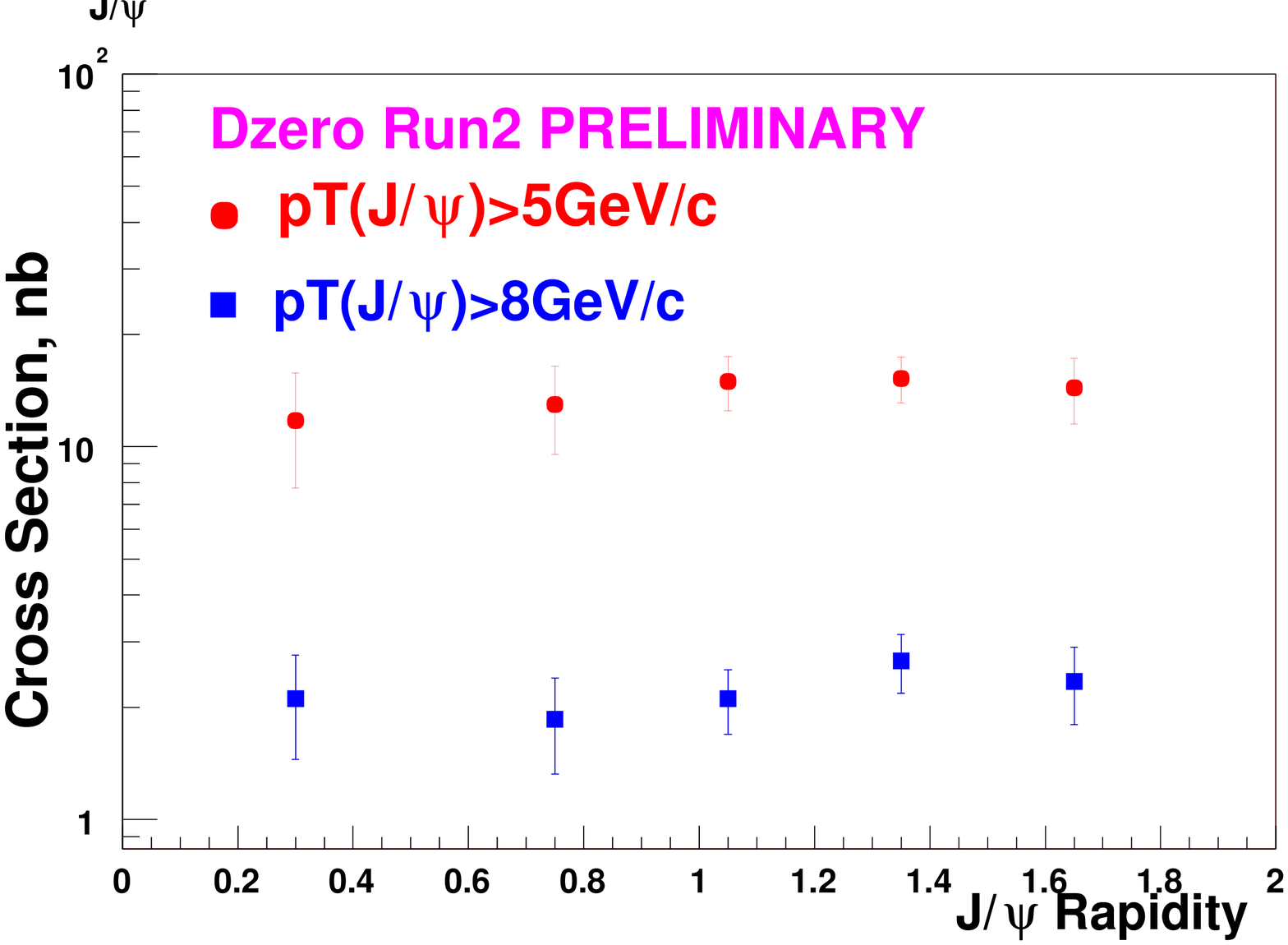,height=1.75in}
\psfig{figure=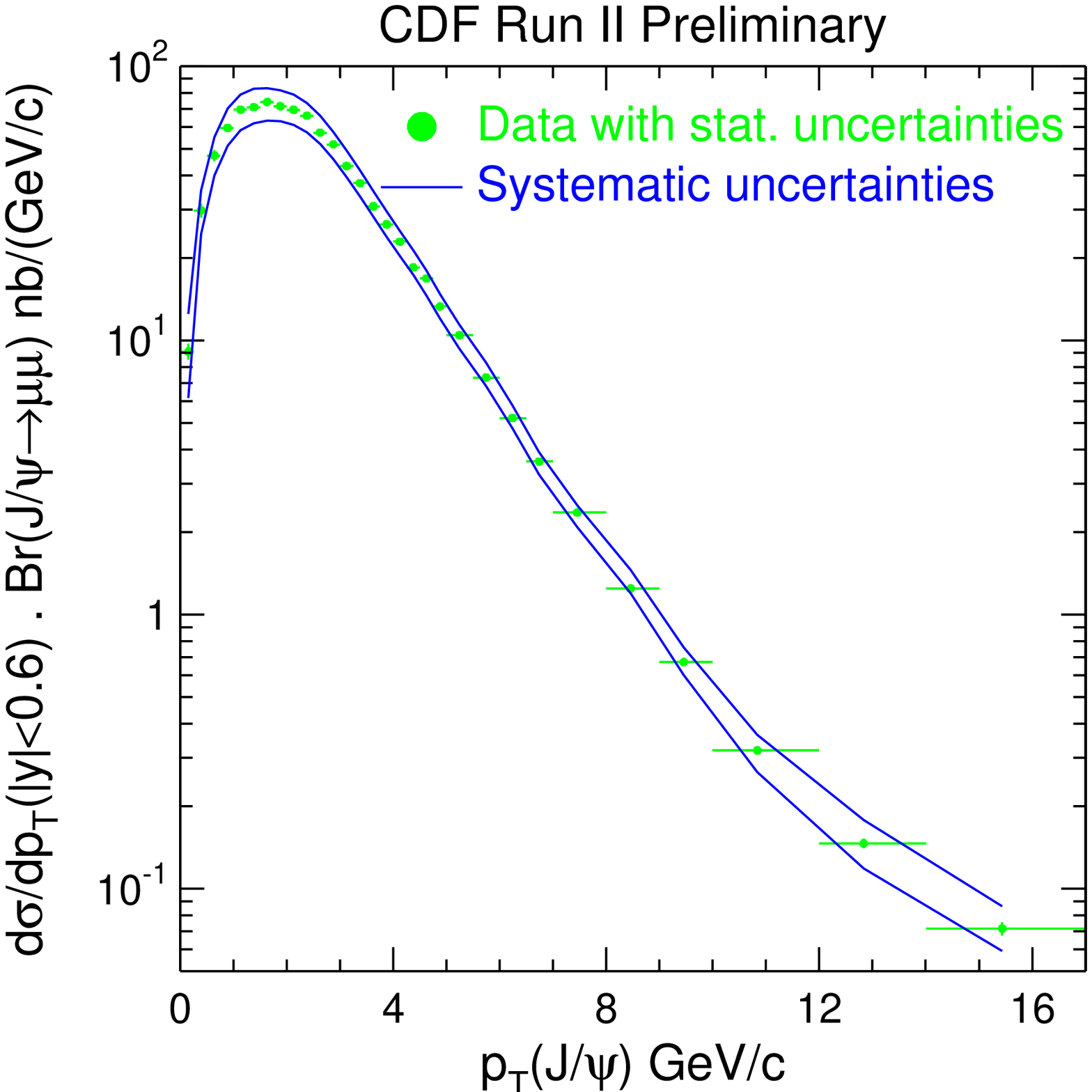,height=2in}
\end{center}
\caption{{\it Left:} The differential $\rm J/\psi$ cross
section in bins of rapidity as measured by \D0 \protect\hfill.\protect\\
\hspace {2cm}{\it Right:} The differential $\rm J/\psi$ cross section
in bins of $p_T$ as measured by CDF\protect\hspace{2.3cm}.\label{fig:jpsid0}}
\end{figure}

\D0 measures the differential $\rm J/\psi$ cross section in bins of
rapidity using $4.74 \rm pb^{-1}$ of data. Using a data sample with an 
integrated luminosity of $39.7 \rm pb^{-1}$ CDF measures the
differential cross section in bins of $p_T$. The resulting
distributions are shown in Fig.~\ref{fig:jpsid0}.

For RUN II the muon trigger momentum thresholds at CDF were lowered to 
$\geq 1.4 \frac{\rm GeV}{\rm c}$ allowing one to trigger on $\rm J/\psi$'s at
rest for the first time. 
The total inclusive cross section has been measured to 
$\sigma(p\bar{p}\ra\rm J/\psi X,|y(\rm J/\psi)|<0.6)\cdot
\mathcal{B}(J/\psi\ra\mu\mu)=240\pm 1 (stat) ^{+35}_{-28} (sys)\rm nb$.

\section{Secondary Vertex Trigger}

With the Silicon Vertex Tracker (SVT)~\cite{SVT} the CDF experiment 
has introduced a novel method to obtain heavy flavor decays.

The SVT uses tracks from the Central Outer Tracking chamber
as seeds to a parallelized pattern recognition in the 
Silicon Vertex Detector. The following linearized track fit returns
track parameters with nearly offline resolution. The precise
measurement of the track impact parameter allows one 
to trigger on displaced tracks from long-lived hadrons containing
heavy flavor. 

Originally designed to select hadronic B-decays the SVT also 
collected a large sample of charm mesons.

\section{Prompt D-meson Cross Section}

The cross section is measured in four fully reconstructed decay modes:
$D^0\rightarrow K^-\pi^+$, $D^{*+}\rightarrow D^0\pi^+$,
$D^+\rightarrow K^-\pi^+\pi^+$ and $D_s^+\rightarrow\phi\pi^+$ using
$5.8 \rm pb^{-1}$ of CDF data.

In order to separate prompt and secondary charm, the impact parameter
distribution of the reconstructed D-meson samples is used.
Mesons originating from B-decay exhibit a large impact parameter.
A fit to the impact parameter distribution yields prompt production 
fractions of $88.6\pm0.4\pm3.5\%$ for $D^0$, 
$88.1\pm1.1\pm3.9\%$ for $D^{*+}$ , $89.1\pm0.4\pm2.8\%$
for $D^+$ and $77.3\pm3.8\pm2.1\%$ for $D_s^+$-mesons averaged over 
all $p_T$ bins. 

\begin{figure}[!h]
\begin{center}
\psfig{figure=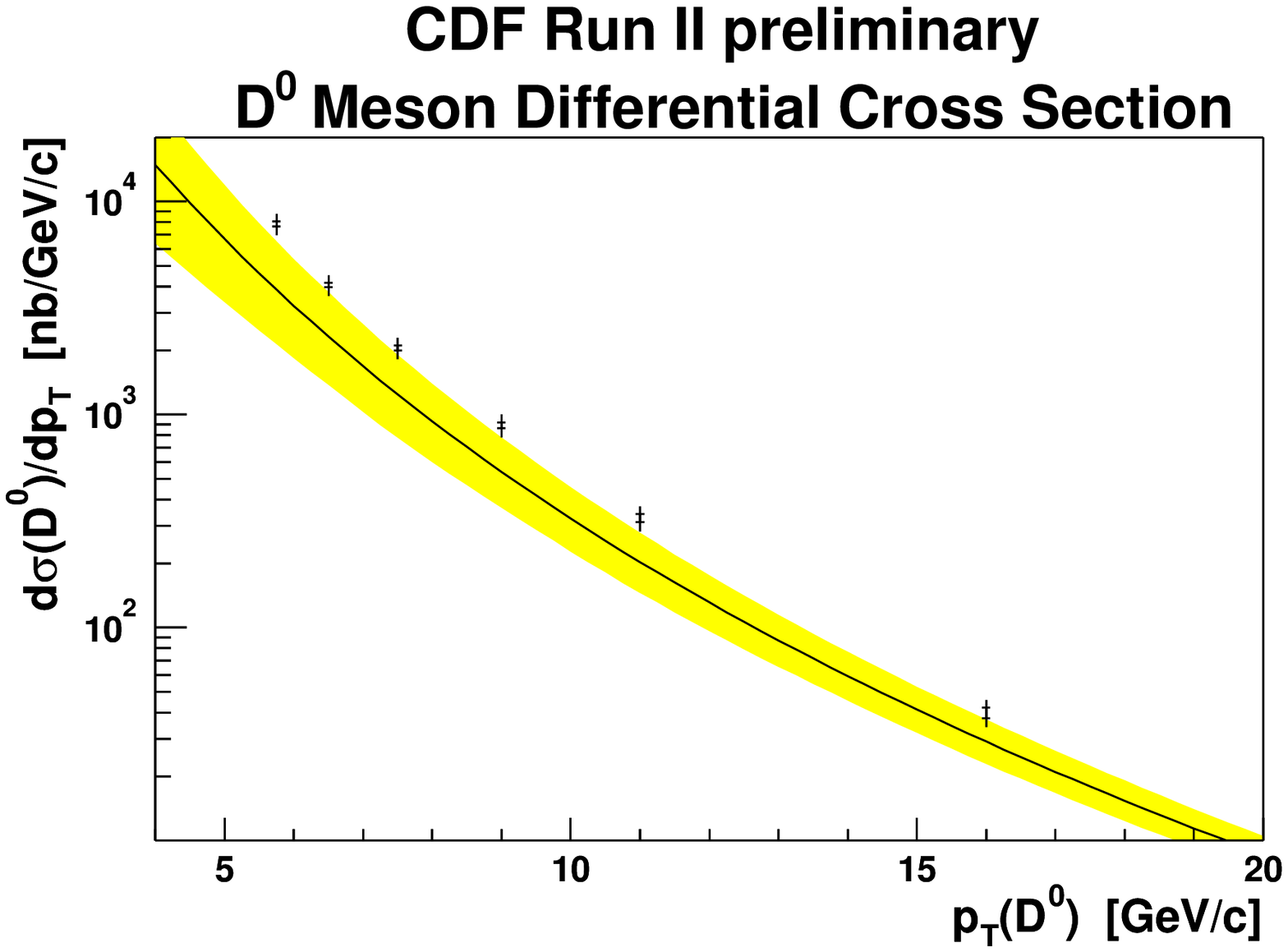,height=1.5in}
\psfig{figure=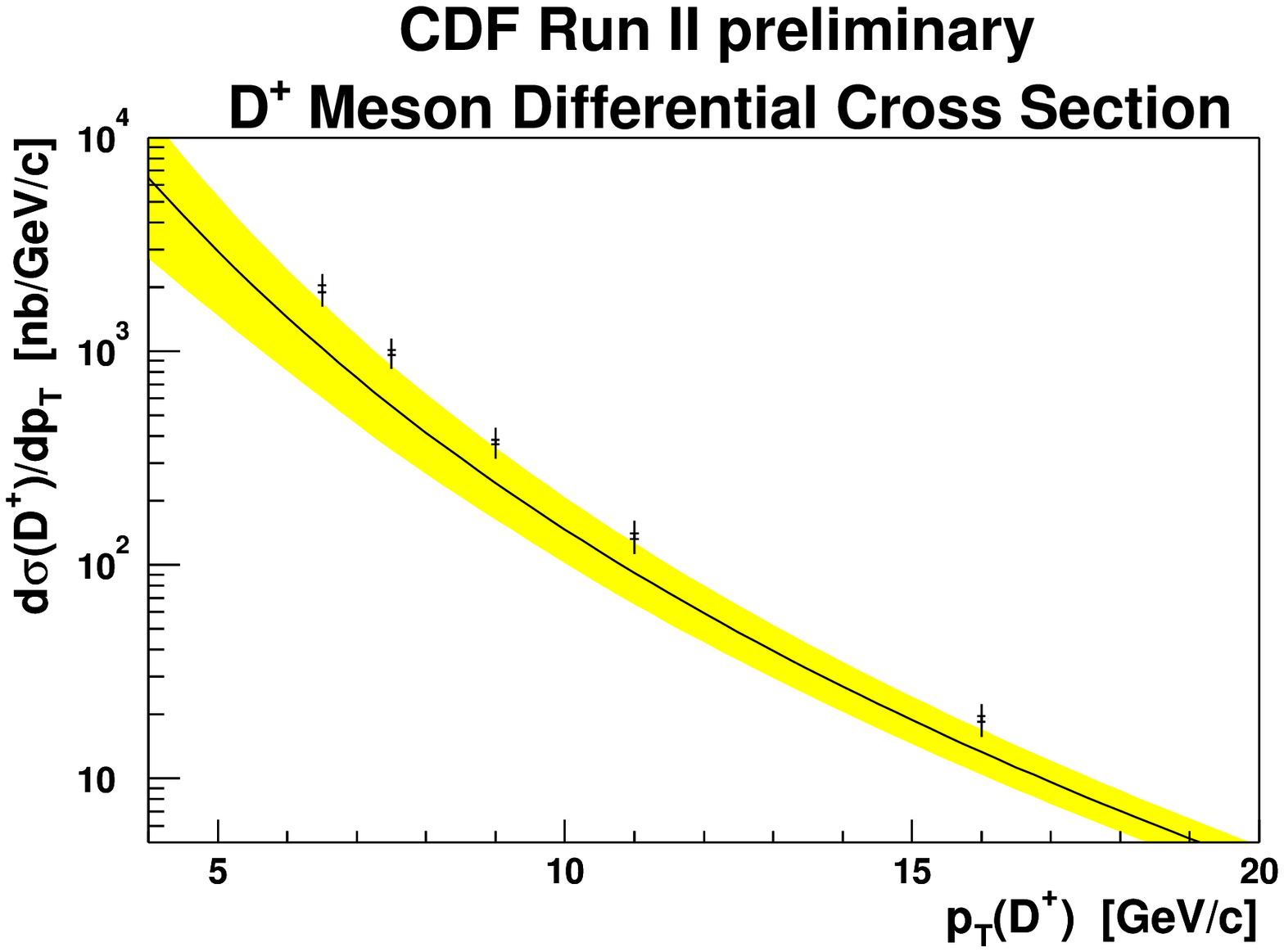,height=1.5in}\\
\psfig{figure=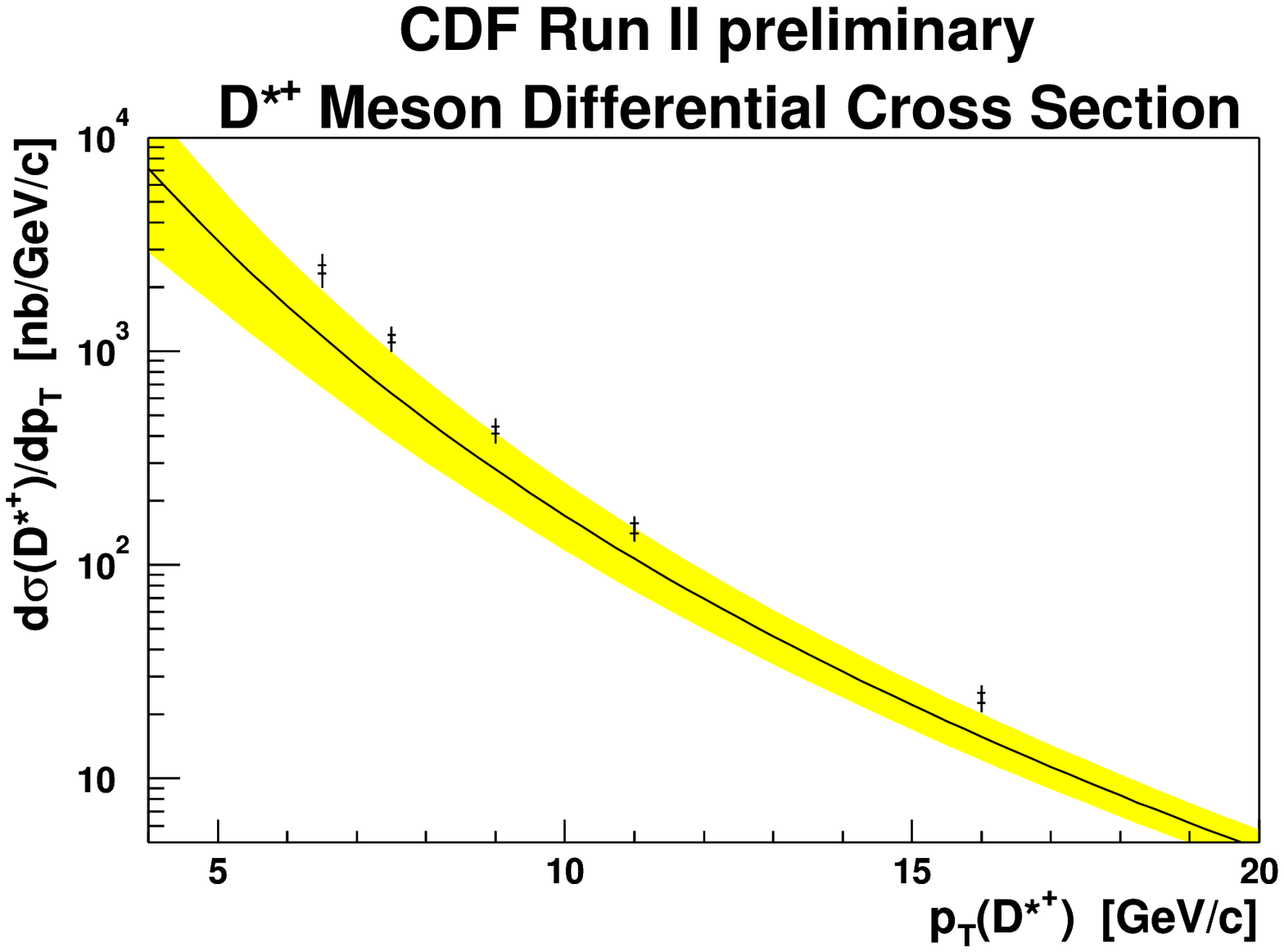,height=1.5in}
\psfig{figure=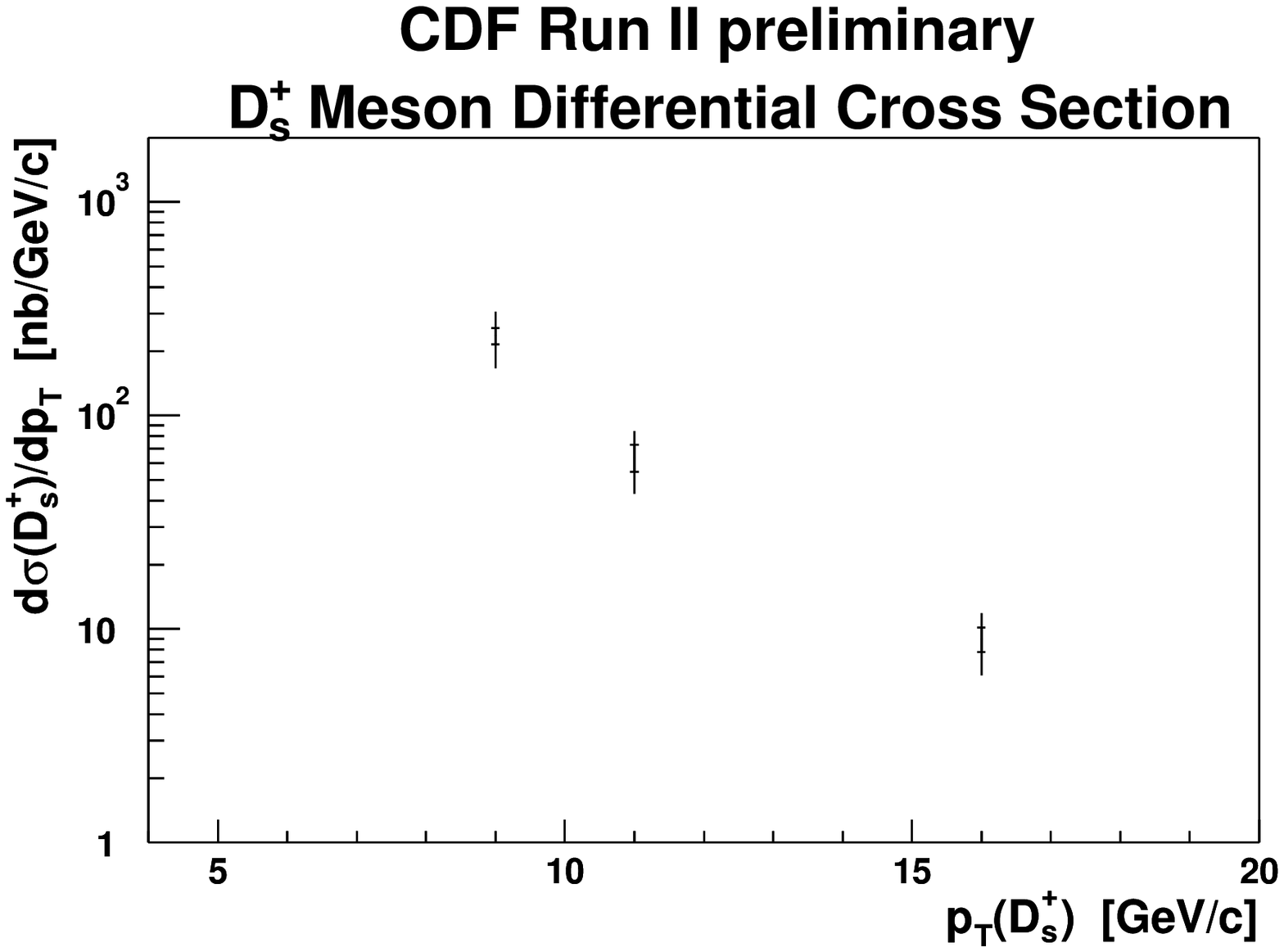,height=1.5in}
\end{center}
\caption{ Differential cross section for the four reconstructed
D-mesons. A theoretical prediction from a Fixed Order Next-to-Leading-Log calculation is overlaid~\protect\cite{DTH}.
\label{fig:dxsection}}
\end{figure}

The measured prompt differential cross section is shown in
Fig.~\ref{fig:dxsection}. The total cross sections are found to be: 
$\sigma(D^0,p_T\geq5.5 \rm \frac{GeV}{c}) = 13.3\pm0.2\pm1.5 \mu\rm b$, 
$\sigma(D^{*+},p_T\geq6.0 \rm \frac{GeV}{c}) = 5.4\pm0.1\pm0.8 \mu\rm b$, 
$\sigma(D^+,p_T\geq6.0 \rm \frac{GeV}{c}) = 4.3\pm0.1\pm0.7 \mu\rm b$, 
$\sigma(D^+_s,p_T\geq8.0 \rm \frac{GeV}{c}) = 0.75\pm0.05\pm0.22 \mu\rm b$.

\section{$\boldmath m_{D^+_s}-m_{D^+}$ Mass Difference}

\begin{figure}[!h]
\begin{center}
\psfig{figure=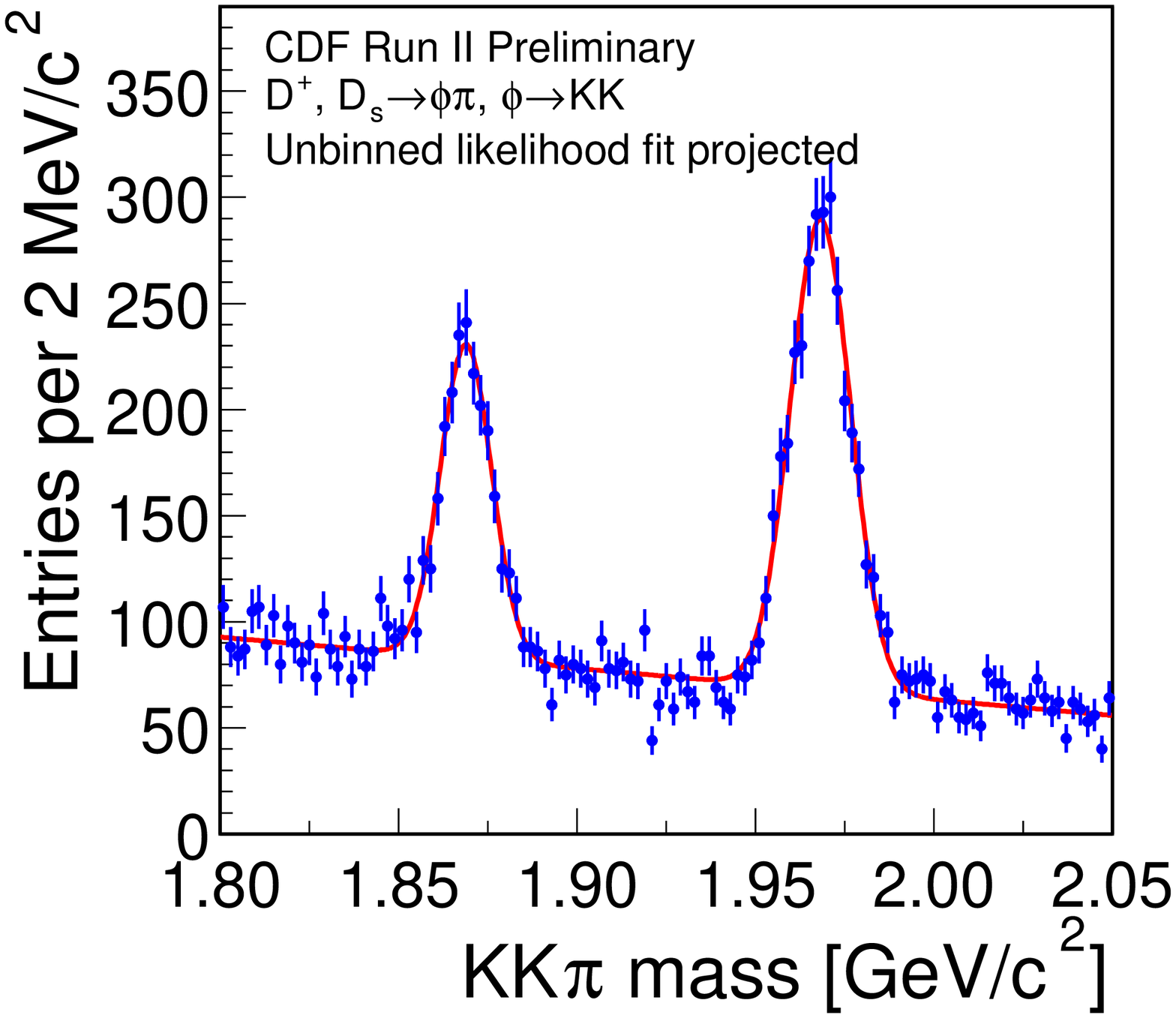,height=2in}
\psfig{figure=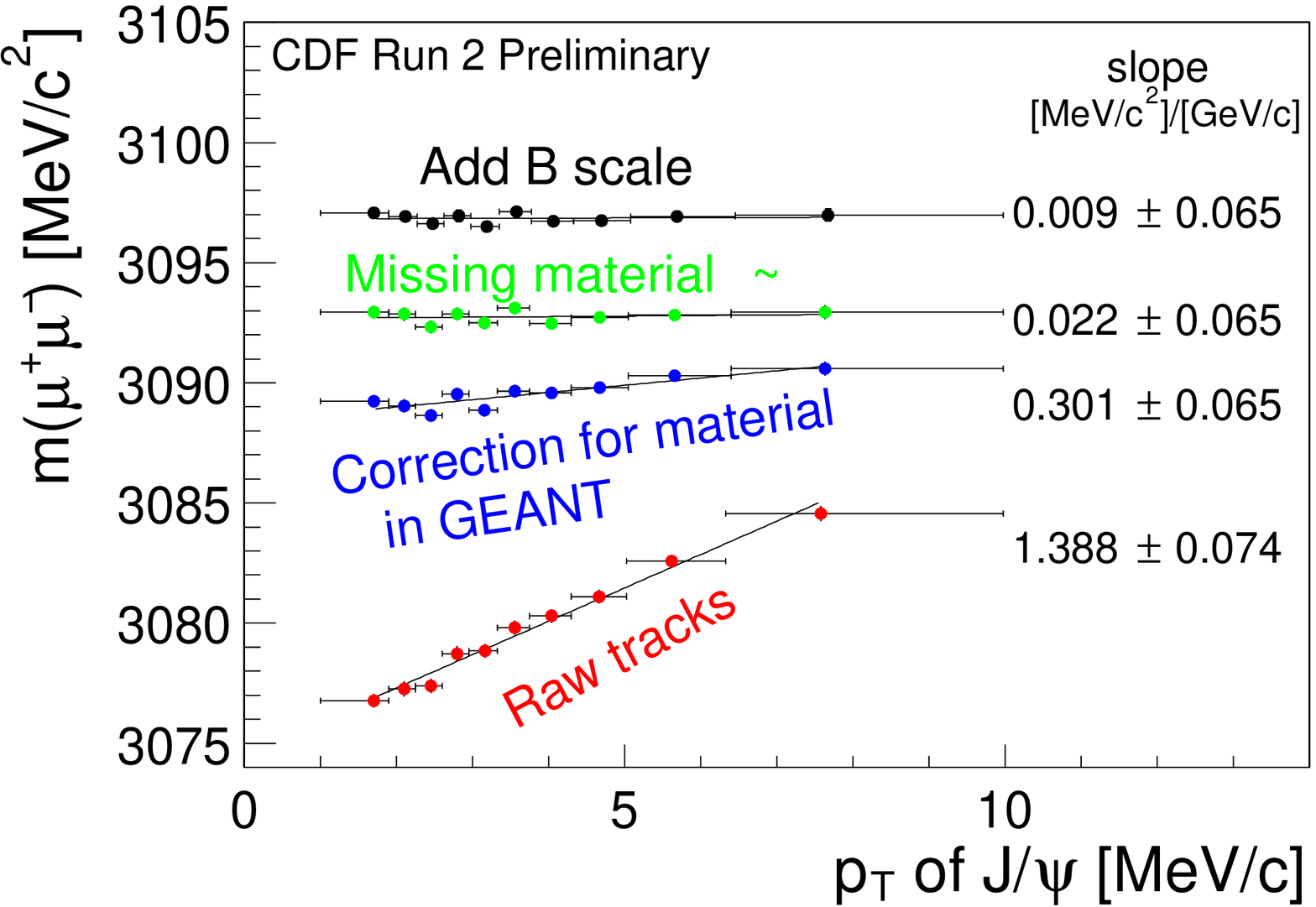,height=2in}
\end{center}
\caption{{\it Left:} Calibration of the momentum scale using the
momentum dependence of the $\rm J/\psi$ mass.\protect\\
{\it Right:} The reconstructed $D_s^+\ra\phi\pi$ and $D^+\ra\phi\pi$
mass distribution. \protect\hspace{0.75cm}\protect\label{fig:ddsmass}}
\end{figure}

The $m_{D^+_s}-m_{D^+}$ mass difference provides a test for HQET and lattice QCD. 
This measurement~\cite{DM}, using only an integrated
luminosity of $11.6 \rm pb^{-1}$, relies on $D^+_s$ and $D^+$ $\ra\phi\pi^+$
decays into $\phi\pi^+$as shown in Fig.~\ref{fig:ddsmass}. 
Using the same decay mode has the advantage of canceling systematics.

For a mass measurement a calibrated momentum scale is the key issue.
A large $\rm J/\psi$ sample was used to calibrated energy loss and 
magnetic field. Slopes in the momentum dependence of the $\rm J/\psi$
are attributed to uncorrected energy loss. The corrections are
adjusted to account for missing material. 
The overall mass shift with respect to the well measured world 
average $\rm J/\psi$ mass is used to fine tune the magnetic field. This is
illustrated in Fig.~\ref{fig:ddsmass}.

The competitive measurement results in $m_{D^+_s}-m_{D^+} =
99.41\pm0.38_{stat}\pm0.21_{sys}\frac{\rm MeV}{\rm c^2}$

\section{Cabbibo Suppressed Decays and CP Violation}

\begin{figure}[!h]
\begin{center}
\psfig{figure=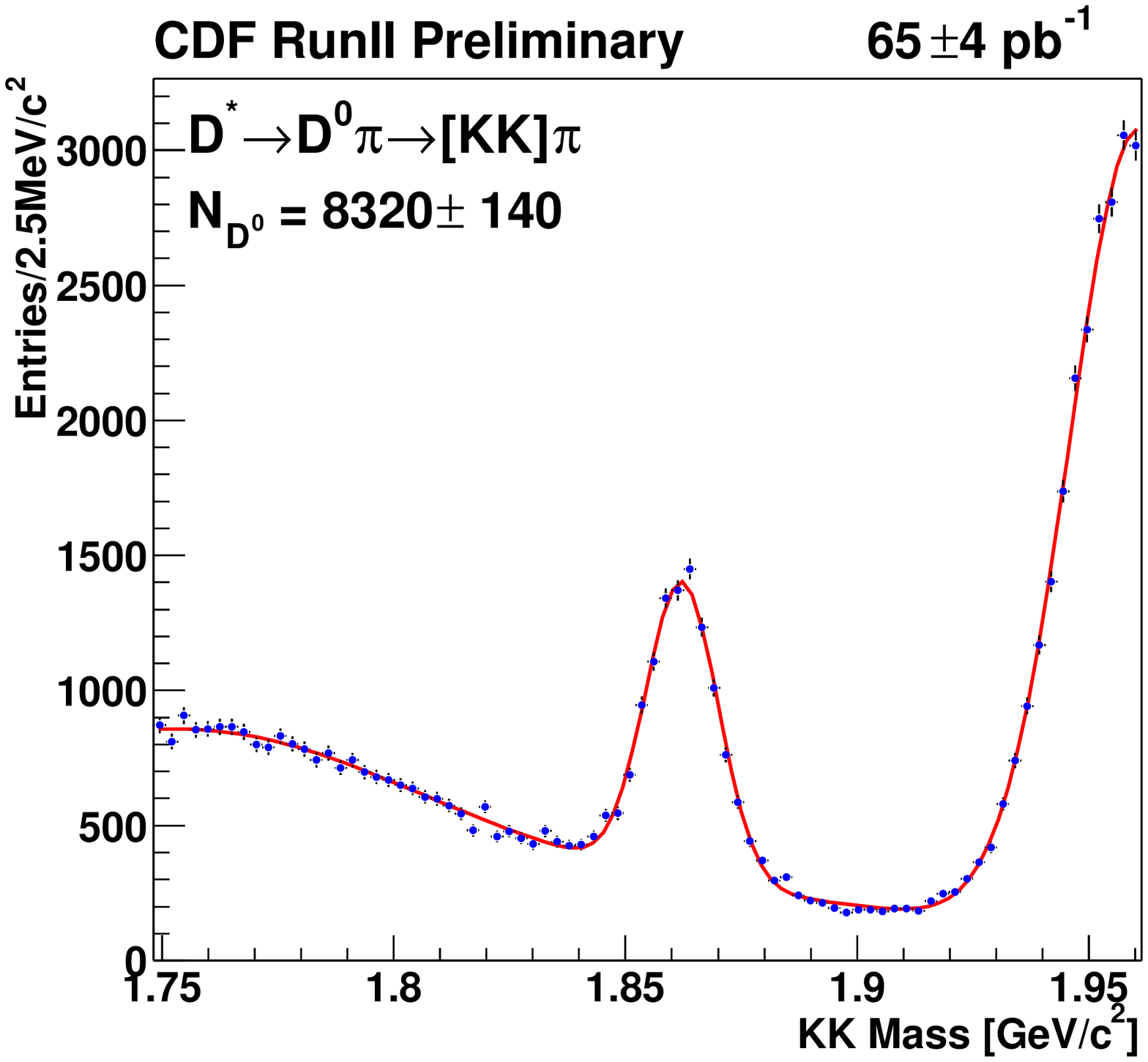,height=1.75in}
\psfig{figure=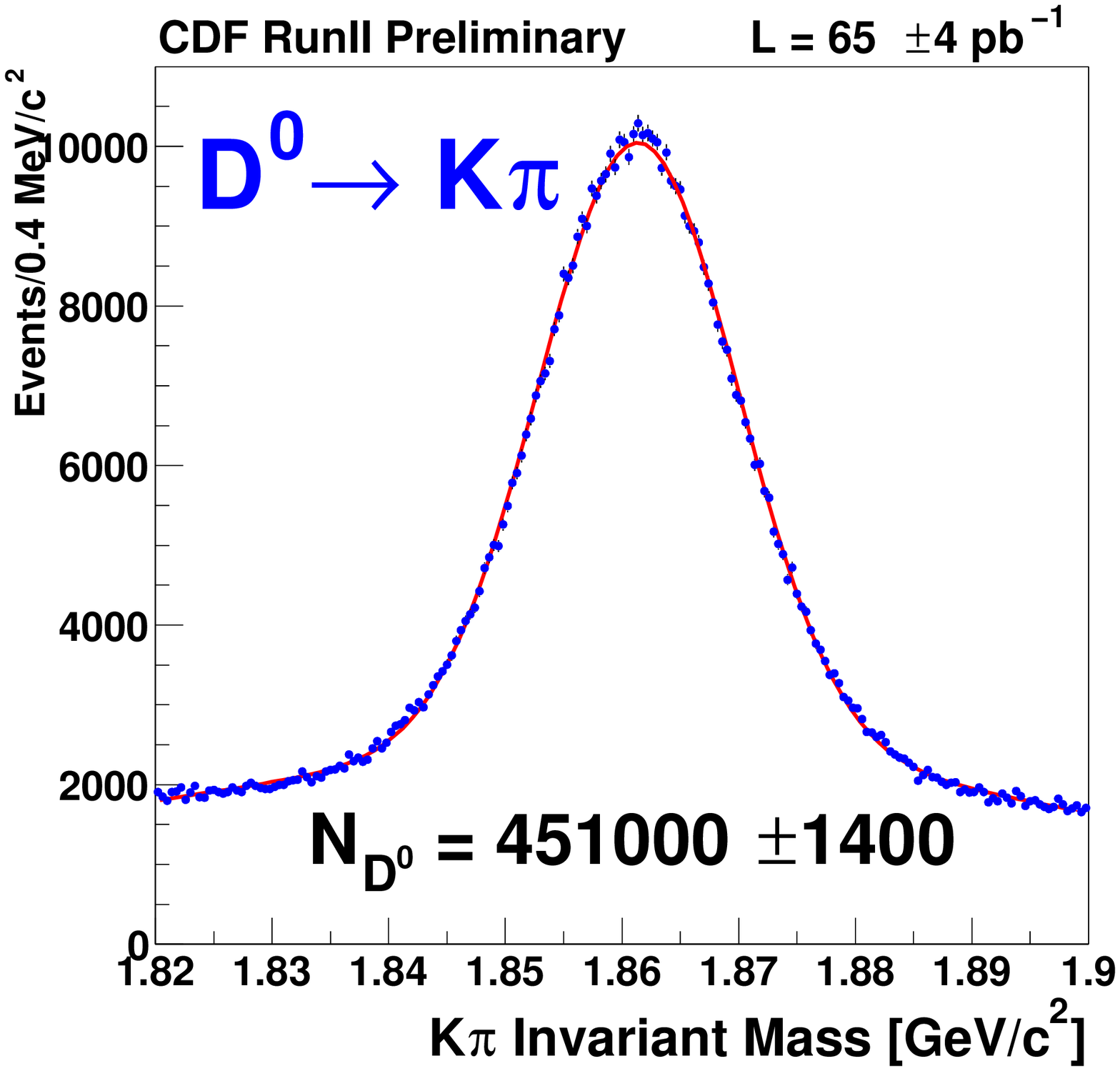,height=1.7in}
\psfig{figure=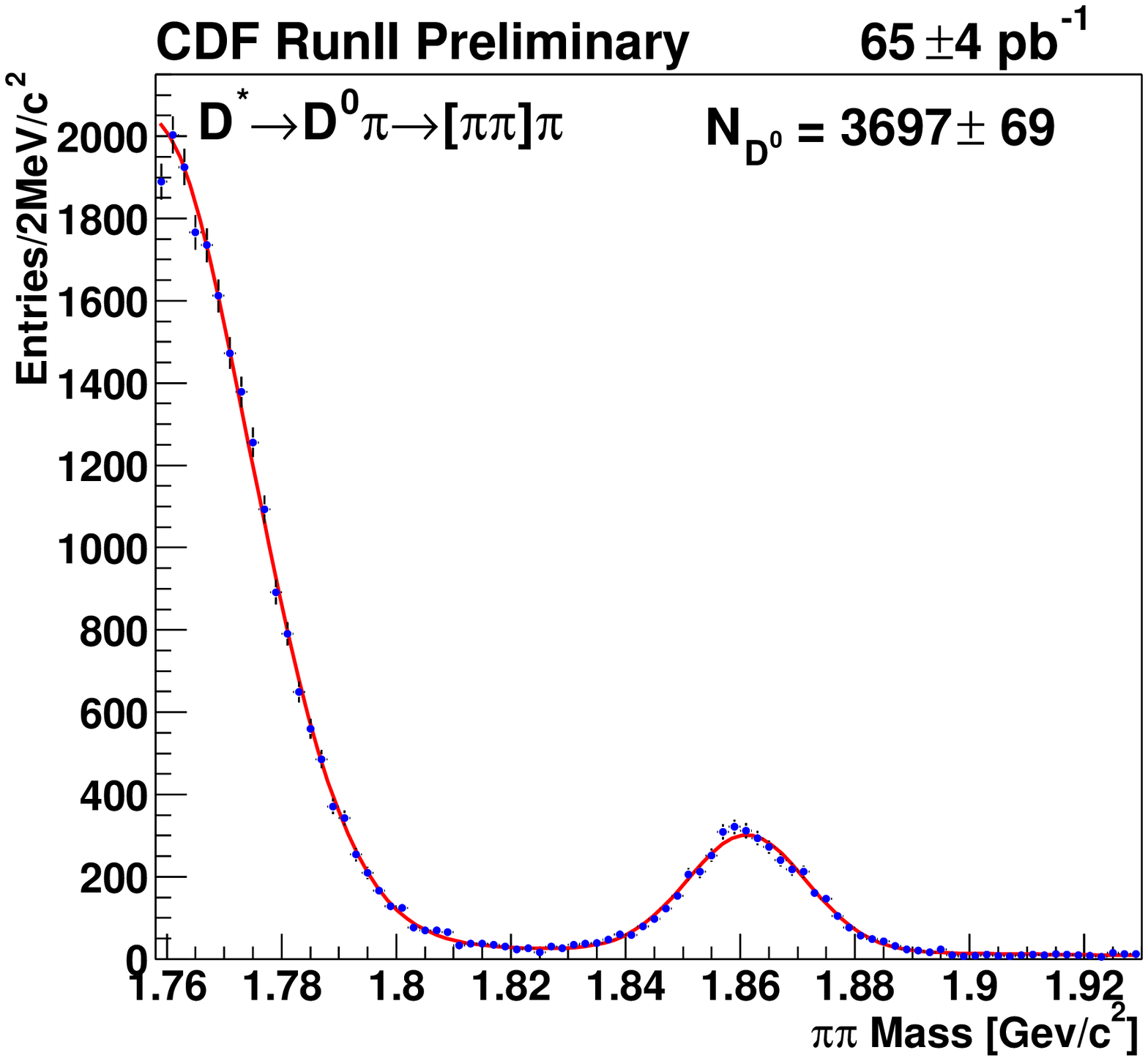,height=1.75in}
\end{center}
\caption{The reconstructed $D^0$ decay modes. 
It can be seen that the reflections are well separated. In order to
determine the flavor of the $D^0$-mesons, they are reconstructed 
from $D^*$ decays.\label{fig:d0br}}
\end{figure}

Utilizing the huge sample of $D^0$-mesons in $65 \rm pb^{-1}$ 
integrated luminosity collected with the secondary
vertex trigger CDF measures the relative branching fractions:
\[
\frac{\Gamma(D^0\ra K^+K^-)}{\Gamma(D^0\ra K^+\pi^-)}=9.38\pm0.18_{stat}\pm0.10_{sys}\%
\]
\[
\,\frac{\Gamma(D^0\ra\pi^+\pi^-)}{\Gamma(D^0\ra K^+\pi^-)}=3.686\pm0.076_{stat}\pm0.036_{sys}\%
\]
This result compares favorably with the current best measurement~\cite{FOCUS}.
The reconstructed decays are shown in Fig.~\ref{fig:d0br}. 
The CP violating decay rate asymmetries $A=\frac{\Gamma(D^0\ra
f)-\Gamma(\bar{D^0}\ra f)}{\Gamma(D^0\ra f)+\Gamma(\bar{D^0}\ra f)}$ are also measured.
It is found that $A(D^0\ra K^+K^-) = 2.0\pm1.7_{stat}\pm0.6_{sys}$ and
$A(D^0\ra\pi^+\pi^-) = 3.0\pm1.9_{stat}\pm0.6_{sys}$, comparable
to previous measurements~\cite{CLEOII}.

\section{Search for the FCNC Decay $\boldmath D^0\rightarrow\mu^+\mu^-$}

A search for the flavor changing neutral current (FCNC) decay 
$D^0\rightarrow\mu^+\mu^-$ has been conducted at CDF based on 
$69 \rm pb^{-1}$ of data.
This branching ratio is $\mathcal{O}(10^{-13})$ in the standard model, 
but can be enhanced up to $\mathcal{B}(D^0\rightarrow\mu^+\mu^-)\sim3.5\times10^{-6}$
in R-parity violating SUSY models. No signal is observed with 1.7
background events expected.
Using the data sample from the secondary vertex trigger provides a well
measured normalization mode $D^0\rightarrow\pi^+\pi^-$. 
After correcting for relative acceptance an upper limit of:
\[
\mathcal{B}(D^0\rightarrow\mu^+\mu^-)\leq 2.4\times10^{-6} {\rm at\,} 90\%CL
\]
is found. This measurement improves the current world best limit~\cite{DMUMU}
of  $4.1\times10^{-6}$.

\section{Conclusion}

A variety of competitive measurements have been performed,
establishing that the experiments at the Tevatron are back online for physics.
The ability to trigger on displaced vertices opens new and exciting
possibilities for charm physics at hadron colliders.

\section*{Acknowledgments}
We would like to acknowledge the work of all the CDF and \D0 collaborators
that made these results possible. 
Thanks to the conference organizers for a very enjoyable and fruitful meeting.

\end{document}